\documentclass[%
 reprint,
 amsmath,amssymb,
 aps,
]{revtex4-2}
\usepackage{comment}
\usepackage{graphicx}% Include figure files
\usepackage{dcolumn}% Align table columns on decimal point
\usepackage{bm}% bold math
\usepackage{hyperref} 
\hypersetup{colorlinks=True, citecolor=cyan}
\usepackage{amsmath}
\usepackage{mathtools}
\usepackage{here}
\usepackage[normalem]{ulem}

\begin{document}

\preprint{APS/123-QED}

\title{Robustness of the avian compass function \\described by radical pair model against biomagnetic noise}% Force line breaks with \\

\author{Tokio Yoshida}
\email{oikot76time@gmail.com}
\author{Masaya Kunimi}%
 \email{kunimi@rs.tus.ac.jp}
\author{Tetsuro Nikuni}%
 \email{nikuni@rs.tus.ac.jp}

\affiliation{%
 Department of Physics, Tokyo University of Science,\\
 1-3 Kagurazaka, Shinjuku, Tokyo 162-8601,Japan}%

\date{\today}% It is always \today, today,

\begin{abstract}
The magnetic sensing mechanism proposed to exist in avian eyes functions as a compass that detects the geomagnetic inclination aiding their migration. This mechanism is modeled by a quantum spin model known as the radical pair (RP) model. This model suggests that information about geomagnetic inclination is transmitted as biochemical signals via spin-selective recombination of the RP. However, as a delicate quantum system in a warm and noisy in vivo environment, the RP model is likely to be subject to unavoidable environmental noise. In this study, we develop a model that incorporates environmental magnetic noise, based on a Lindblad-type master equation. Our model includes the type of biological tissue and the temperature of the biological system as parameters, accounting for their effects. The results of numerical calculations indicate that the compass function of the RP model is robust against intrinsic magnetic noise arising from thermal fluctuations in the electromagnetic field within realistic biomagnetic range. However, noise from an environmental magnetic field far exceeding the biological magnetic range, on the order of hundreds to thousands of $\mu$T, significantly affects and disrupts the compass function. This work provides a foundation for quantifying environmental influences on open quantum systems in biological settings and offers a tool for connecting experimental measurements of environmental magnetic fields with quantum noise modeling.
\end{abstract}

\maketitle

%\tableofcontents

\section{\label{sec:intro}Introduction}

Magnetoreception observed in certain living systems is one of the key topics in quantum biology \cite{PRL, 3theme}. This ability allows organisms to sense the Earth's weak magnetic field, referred to as geomagnetism, and use it for navigation. For example, the European robin, a migratory bird, benefits from this magnetoreception mechanism to travel long distances to its wintering grounds \cite{Ritz}. The theory developed to explain the mechanism behind the magnetoreception of these migratory birds is known as the radical pair (RP) model.

By the 1950s, behavioral experiments confirmed that migratory birds such as the European robin can detect geomagnetic field and use this ability to navigate \cite{BE,Ritz,frank,light2009}. Based on these findings, the RP model was proposed \cite{Schulten1978}. This model suggests that the spin-selective recombination reactions of RP \cite{Haberkorn}, which are influenced by anisotropic interactions of the geomagnetic field, yield products whose quantities depend on the angle of the inclination of the geomagnetic field. These angle-dependent products act as biochemical signals, providing birds
with directional information about the geomagnetic field. Detailed explanations of this recombination process, the RP generative process, and the molecular and chemical properties of RP are given in Refs. \cite{PRL,relativeorientation,Luo2024}.

Since the initial proposal of the RP model, various adaptations have been developed, making it an interdisciplinary research focus in biochemistry, quantum chemistry, and quantum information science. These studies have explored how the structure of RP, intrinsic interactions, and factors such as entanglement and coherence during recombination reaction processes contribute to its function as a magnetic compass \cite{quadrupole, magnetization, Bell, pauls,superoxide}.

However, this system operates in a noisy, warm, and wet in vivo environment raising the question of whether the delicate chemical compass can function effectively under these conditions. In studies aimed at answering this question, the RP model has been formulated within the context of open quantum systems, and numerical simulations incorporating the effect of noise have been performed \cite{PRL,envinduced,relativeorientation,WaterMP,magnetization}. These studies have explored how environmental noise affects the functionality of the chemical compass. In these studies, various perturbations to spin states have been introduced to represent noise. However, further investigation is needed for a more detailed description of the measurable and specific biological environments surrounding the RP.

%However, some of these models lack detailed descriptions of the measurable, specific biological environments surrounding the RP.

The aim of the present paper is to clarify the effects of biomagnetism on the avian chemical compass described by the RP model. To achieve this, we propose a modified RP model that incorporates the environmental noise originating from the magnetic fields in the medium surrounding the RP. Using this model, we calculate the angle dependence of the singlet yield from the RP to investigate the robustness of the compass function against the noise. The noise intensity in our model is characterized by the coupling constants that describe the strength of the interaction between the RP and the environment. The coupling constants can be determined by factors of the biological environment, such as temperature and the type of medium (biological tissues). From these coupling constants, indicative quantities characterizing the environmental condition, such as the magnitude of environmental magnetic field, can be estimated. By comparing this magnetic field with the biological magnetism measured in actual living systems, we discuss whether a chemical compass can function in real biological magnetic fields.

This paper is organized as follows. In Sec.~\ref{subsec:RPwithoutE}, we describe the details of the RP model. In Sec.~\ref{subsec:Noise}, we present the master equation used for the numerical calculations and discuss the relationship between the properties of the RP system and the coupling constants representing the noise intensity from the environment. In Sec.~\ref{subsec:CC}, we  illustrate the environmental information encoded within these coupling constants. In Sec.~\ref{subsec:Biomagnetism}, we show how the model is adapted to realistic biomagnetic environments. The numerical results are shown in Sec. \ref{sec:result}. In Sec.~\ref{sec:Discussion}, we discuss the effects of the biological magnetic environment on the RP model. Finally, the conclusions of this paper are summarized in Sec.~\ref{sec:summary}.

\section{\label{sec:model}Model}

In this section, we begin with Sec.~\ref{subsec:RPwithoutE} by reviewing the conventional RP model to provide a complete background and discussing its compass functionality. In Sec.~\ref{subsec:Noise}, we incorporate environmental noise, which arises from magnetic fields propagating through the biological medium, into the RP model. In Sec.~\ref{subsec:CC}, we present the coupling constants that determine the noise intensity, adjusting them to reflect realistic biological environmental conditions. Finally in Sec.~\ref{subsec:Biomagnetism}, we propose a formula
that relates the coupling constants of the noise to the environmental magnetic field.

\subsection{\label{subsec:RPwithoutE}RP model without environmental noise}
The model described in this subsection is mainly based on \cite{PRL}.
We consider a minimal model for the RP \cite{PRL, envinduced, Kominis}. This model consists of one of the nuclei in the RP and the unpaired electrons within each radical:
\begin{align*}\label{eq:hamitonian}
    \hat{H}(\vartheta)&=\mathbf{\hat{I}} \cdot \mathbf{A} \cdot \mathbf{\hat{S}}_{1}+\gamma \mathbf{B}_{\mathrm{GM}}(\vartheta) \cdot\left(\mathbf{\hat{S}}_{1}+\mathbf{\hat{S}}_{2}\right)\\
    &=\sum_{\mu=x,y,z}A_{\mu}\hat{I}_{\mu}\hat{S}_{1\mu}+\gamma \sum_{i=1,2}\sum_{\mu=x,y,z}B_{\mathrm{GM}}(\vartheta)_{\mu}\hat{S}_{i\mu},
    \stepcounter{equation}\tag{\theequation}
\end{align*}
where $\hat{\mathbf{I}}$ is the spin-1/2 operator for the nucleus, while $\hat{\mathbf{S}}_1$ and $ \hat{\mathbf{S}}_2$ are the spin-1/2 operators for the unpaired electrons. Following the notation in Ref. \cite{PRL}, we refer to the Pauli operators as spin operators, omitting the usual factor of 1/2. Specifically, the spin operators are defined as $\hat{\bf{I}}=(\hat{\sigma}_x,\hat{\sigma}_y,\hat{\sigma}_z)_{\rm{I}}$ for the nucleus and  $\hat{\bf{S}}_i=(\hat{\sigma}_x,\hat{\sigma}_y,\hat{\sigma}_z)_{i}$ for the unpaired electrons ($i = 1, 2$). Here, the Pauli matrices include, for example, $\hat{\sigma}_z=\operatorname{diag}(1,-1)$. This RP Hamiltonian is defined within the Hilbert space $\mathcal{H}_{\mathrm{RP}}$, which spans the spin degrees of freedom of the RP. Based on a previous study \cite{Ritz}, one of the unpaired electrons is assumed to be uncoupled to the nucleus. The quantity $\mathbf{A}$ is a $3\times 3$ tensor representing the anisotropic hyperfine interaction (HFI) coupling between the nucleus and one of the unpaired electrons. We assume a simple diagonal anisotropy: $A_{xx}=A_{yy}=\frac{1}{2}A_{zz}$, with $A_{zz}=10^{-5}$ meV \cite{PRL}. The second term of Eq.~(\ref{eq:hamitonian}) represents the Zeeman coupling between the geomagnetic field and the RP. The constant $\gamma\equiv \frac{1}{2}\mu_{\mathrm{B}} g$ is the gyromagnetic ratio, $\mu_{\mathrm{B}}$ is the Bohr magnetron, and $g$ is the electron spin g-factor, taken as 2.
The geomagnetic field vector $\bm{B}_{\rm GM}(\vartheta)$ is given by
\begin{align}
    \mathbf{B}_{\mathrm{GM}}(\vartheta)\equiv\mathbf{B}_{\mathrm{GM}}(\varphi,\vartheta)|_{\varphi=0}= B_0\left.\begin{pmatrix}
        \cos \varphi \sin \vartheta \\ \sin\varphi \sin \vartheta \\ \cos \vartheta
    \end{pmatrix}\right|_{\varphi=0}.
\end{align}
 Here, the constant of the geomagnetic field $B_0$ is set to 47 $\mu$T \cite{Ritz}. For simplicity, we align the $z$-axis in spherical coordinates with the horizontal plane and assume that the azimuthal angle $\varphi$ is zero without loss of generality, as the HFI tensor is rotationally symmetric around the $z$-axis. Additionally, $\vartheta$ represents the angle of inclination of the geomagnetic field to be detected by the avian. 

To fully describe the recombination process of the RP, it is necessary to consider more than just the spin degree of freedom. Specifically, spin degrees of freedom alone cannot account for the types of product generated when the RP recombines through the singlet or triplet channels (i.e., singlet or triplet products). To capture this aspect, we introduce a shelving state, represented by an additional subspace $\mathcal{H}_{\mathrm{shelf}}$. This augmented Hilbert space, $\mathcal{H}_{\rm{rec}}$, allows us to fully describe the recombination process, and is defined as a direct sum:
 \begin{align}
\mathcal{H}_{\rm{rec}}=\mathcal{H}_{\rm{RP}}\oplus\mathcal{H}_{\mathrm{shelf}}.
 \end{align}
 The subspace $\mathcal{H}_{\mathrm{shelf}}$ is spanned by two shelving states, $|S\rangle$ and $|T\rangle$, which correspond to the molecular products generated via the singlet and triplet recombination channels, respectively. These shelving states form an orthonormal basis. In this extended Hilbert space, the density operator for the RP and the recombination process $\hat{\rho}(t)$ is expressed as
 \begin{align}
     \hat{\rho}(t)=\left(\begin{array}{@{}c|c@{}}
 \hat{\rho}_{\rm{spin}}(t)
& \mbox{\Large 0} \\
\hline
{\mbox{\Large 0}} &
\begin{matrix}
\rho_{\rm{S}}(t) & 0 \\
0 & \rho_{\rm{T}}(t)
\end{matrix}
\end{array}\right).
 \end{align}
 Here, $\hat{\rho}_{\mathrm{spin}}(t)$ denotes the density operator describing the spin state of the RP. The lower right $2\times2$ block represents the shelving subspace. Once recombination is complete and the system reaches a steady state at $t_{\mathrm{fin}}$, the populations of $\rho_{\rm{S}}(t_{\mathrm{fin}}) $ and $\rho_{\rm{T}}(t_{\mathrm{fin}})$ correspond to the fractions of generated singlet and triplet products, respectively. These fractions are referred to as the singlet yield and the triplet yield, and the satisfy $\rho_{\rm{S}}(t_{\mathrm{fin}})+\rho_{\rm{T}}(t_{\mathrm{fin}})=1$. In this scheme, the off-diagonal elements between the spin state and the shelving state, i.e., the upper right and lower left blocks in Eq. (4), are approximated to be zero, and the transitions between the spin state and shelving state are treated perturbatively.

In the initial state, the populations of the shelving states are assumed to be zero. The nonzero populations of $\rho_{\rm{S}}(t) $ and $\rho_{\rm{T}}(t)$ arise solely from the action of recombination transition operators, which are defined as
 \begin{align}
    \hat{T}_{T_{k},\downarrow}&=|T\rangle\langle t_{k},\downarrow|, \\
    \hat{T}_{S,\uparrow}&=|S\rangle\langle s,\uparrow|.
\end{align}
 Here, $\langle\uparrow|$ and $\langle \downarrow|$ denote nuclear spin states, while $\langle t_k|$, $\langle s|$ represent the states of the unpaired electron pair, which are defined as follows. The singlet state is
 \begin{equation}
    |s\rangle=\frac{1}{\sqrt{2}}\left(|\uparrow\rangle_{1}|\downarrow\rangle_{2}-|\downarrow\rangle_{1}|\uparrow\rangle_{2}\right),
\end{equation}
and the triplet states $|t_k\rangle$ are 
\begin{align}
    |t_+\rangle&= |\uparrow\rangle_{1}|\uparrow\rangle_{2}, \\
    |t_0\rangle&=\frac{1}{\sqrt{2}}\left(|\uparrow\rangle_{1}|\downarrow\rangle_{2}+|\downarrow\rangle_{1}|\uparrow\rangle_{2}\right),\\
    |t_-\rangle&=|\downarrow\rangle_{1}|\downarrow\rangle_{2}.
\end{align}
The indices $1,2$ correspond to the two unpaired electrons.

To describe the dynamics of this RP system, we use a Lindblad-type master equation \cite{theoryOQS,QP}:
\begin{align*}\label{eq:LMEwithoutE}
    \frac{\mathrm{d} \hat{\rho}(t)}{\mathrm{d} t}&\equiv\mathcal{M}_{\mathrm{RP}}[\hat{\rho}(t)] \\
    &\equiv -\frac{i}{\hbar}[\hat{H}(\vartheta), \hat{\rho}(t)]+k_{\mathrm{rec}} \sum_{k}\mathcal{L}_{\hat{T}_k}[\hat{\rho}(t)],
    \stepcounter{equation}\tag{\theequation} 
\end{align*}
where $k_{\rm rec}$ represents a recombination rate. In this paper, we assume that $k_{\rm rec}=10^6{\,\rm s}^{-1}$ for all recombination processes, for simplicity \cite{PRL,envinduced}. This value corresponds to $1\mu{\rm s}$ lifetime of the RP, which is based on behavioral experiments and previous studies ~\cite{Ritz, PRL}.
We define $\mathcal{L}_{\hat{P}}$ as the Lindblad-type superoperator
\begin{align*}\label{eq:sup}
    \mathcal{L}_{\hat{P}}[\hat{\rho}(t)]&=\hat{P} \hat{\rho}(t) \hat{P}^{\dagger}-\frac{1}{2}\Big\lbrack\hat{P}^{\dagger} \hat{P} \hat{\rho}(t)+\hat{\rho}(t) \hat{P}^{\dagger}\hat{P}\Big\rbrack,
    \stepcounter{equation}\tag{\theequation} 
\end{align*}
where $\hat{P}$ represents a general Lindblad operators.

The action of the transition operators $\hat{T}_k$ in (\ref{eq:LMEwithoutE}) leads to a decrease in the trace of the spin part of the density operator $\hat{\rho}_{\mathrm{spin}}(t)$ due to recombination. However, this decrease  is compensated by the accumulation in the populations of the shelving states. Consequently, the trace of the total density operator in (\ref{eq:LMEwithoutE}) remains preserved, ensuring $\operatorname{Tr}\hat{\rho}(t)=1$.

Furthermore, we assume that the spin state of the unpaired electrons in the RP is initially in a fully mixed singlet state, given by $\hat{\rho}(0)=\frac{1}{2}(|s,\uparrow\rangle\langle s,\uparrow|+|s,\downarrow\rangle\langle s,\downarrow|)$ \cite{PRL,initialS}. 

\begin{figure}[t]
    \centering
    \includegraphics[width=7cm,clip]{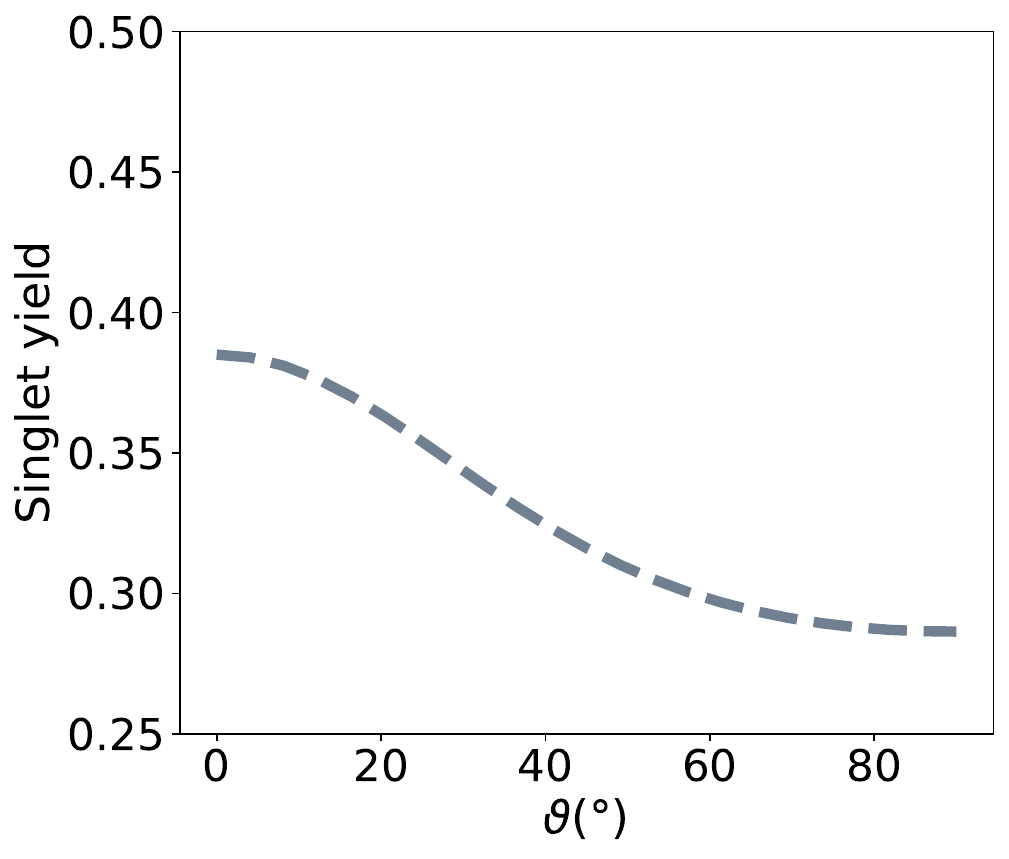}
    \caption{The angular dependence of the singlet yield in the RP model without environmental noise. This yield curve serves as a reference for evaluating the compass function of the RP model.}
    \label{fig:RPM}
    \vspace{-0.75em}
\end{figure}%

By varying the angle of geomagnetic inclination $\vartheta$ and solving the master equation (\ref{eq:LMEwithoutE}) for each value of $vartheta$ \cite{vectorize}, we obtain the population of the singlet shelving state $\rho_{\rm{S}}(t_{\mathrm{fin}})$(i.e., the singlet yield). For completeness, we reproduce the numerical results of the singlet yield as a function of $\vartheta$ reported in Ref.~\cite{PRL}. The results are shown in Fig.~\ref{fig:RPM}. This clearly demonstrates a monotonic dependence on the angle $\vartheta$. Additionally, there is a significant contrast, defined as the difference between the maximum and minimum values of the singlet yield indicating that the singlet yield value can uniquely determine the angle, thereby demonstrating the compass functionality.

\subsection{\label{subsec:Noise}Environmental magnetic noise}

The master equation (\ref{eq:LMEwithoutE}) presented in the previous section does not account for the effects of coupling with the environment. To incorporate these effects, we introduce the Lindblad dissipation terms, which describe the interaction between the RP system and the environment, into the master equation (\ref{eq:LMEwithoutE}):
\begin{align*}
\label{eq:magLME}
    \frac{\mathrm{d} \hat{\rho}(t)}{\mathrm{d} t}&=\mathcal{M}_{\mathrm{RP}}[\hat{\rho}(t)]\\
    &+\sum_{i=1,2}\sum_{\alpha=x,y,z}\sum_{\omega}\gamma_{\mathrm{med}}(\omega,T,Y)\mathcal{L}_{\hat{A}_{i\alpha}(\omega)}[\hat{\rho}(t)],
    \stepcounter{equation}\tag{\theequation} 
\end{align*}
where the superoperator $\mathcal{L}$ has been defined in Eq. (\ref{eq:sup}).
Here, the coupling constant $\gamma_{\mathrm{med}}(\omega,T,Y)\equiv Y\gamma_{\mathrm{med}}(\omega,T)$ represents the strength of the interaction between the system and the environment at a temperature $T$, where we have introduced the parameter $Y$ as a scaling factor to adjust the noise intensity.
In this study, we assume that the coupling between the RP and the environment is mediated by the interaction between the unpaired electron pair and the magnetic field in the medium. In this case, the Lindblad operator $\hat{A}_{i\alpha}(\omega)$ is given by
\begin{align}\label{eq:LO}
    \hat{A}_{i\alpha}(\omega)&\equiv\sum_{\epsilon_m-\epsilon_n=\hbar\omega}\hat{\Pi}_m\hat{\sigma}_{i\alpha}\hat{\Pi}_n,
\end{align}
where $\hat{\Pi}_n\equiv|\phi_n\rangle\langle\phi_n|$ is the projection operator onto the eigenstates $|\phi_n\rangle$ of the RP Hamiltonian $\hat{H}(\vartheta)$ with the eigenenergies $\epsilon_n$ \cite{theoryOQS}. These eigenstates and eigenenergies depend on the geomagnetic field dip angle $\vartheta$. The quantity $\omega$ represents the frequency difference between the eigenstates $|\phi_n\rangle$ and $|\phi_m\rangle$. The summation in Eq. (\ref{eq:LO}) is taken over all pairs of eigenstates with a particular energy difference $\hbar\omega$. In addition, the Pauli operators $\hat{\sigma}_{i\alpha}\;(\alpha=x,y,z)$ act as noise operators, introducing perturbations to each unpaired electron $i=1,2$. 

The function $\gamma_{\mathrm{med}}(\omega,T)$ is the coupling coefficient that encapsulates the properties of the environment and is formally given by \cite{theoryOQS} 
\begin{align}\label{eq:Gamma1}
    \gamma_{\mathrm{med}}(\omega,T)=\frac{\mu_{\rm{B}}^2}{\hbar^2}\int^{\infty}_{-\infty}e^{i\omega t}\Gamma(t)dt,
\end{align}
\begin{align}\label{eq:Gamma2}
    \Gamma(t)=\frac{1}{3}\langle \hat{\mathbf{B}}_{\mathrm{med}}(t)\cdot \hat{\mathbf{B}}_{\mathrm{med}}(0)\rangle.
\end{align}
Here, $\Gamma(t)$ represents the autocorrelation function of the magnetic field in the medium $\hat{\mathbf{B}}_{\mathrm{med}}(t)$, assuming that the medium is isotropic. The operator $\hat{\mathbf{B}}_{\rm{med}}(t)$ represents the fluctuations of the environmental magnetic field, and its average value is assumed to be zero, i.e., $\langle \hat{\mathbf{B}}_{\rm{med}}(t)\rangle=0$. The explicit expression for the coupling coefficient can be obtained by quantizing the electromagnetic field in the medium \cite{CC1, CC2, CC3}. In particular, the explicit expression for the correlation function of the electric field in the medium has been derived in Ref. \cite{CC3}. An analogous correlation function for the magnetic field can also be obtained, and by using this, the explicit form of $\gamma_{\mathrm{med}}(\omega,T)$ is given by the following expression:
\begin{align*}\label{eq:CC}
    &\gamma_{\mathrm{med}}(\omega,T)=\frac{\mu_0\mu_{\mathrm{B}}^2}{3\pi \hbar c^2}\omega^2\lbrack 1+N(\omega,T)\rbrack\\
    &\quad \quad\times\left(\frac{\mathrm{Im}[\epsilon_r^{\mathrm{med}}(\omega)]}{R}+\frac{\omega}{c}\mathrm{Re}\lbrace[\epsilon_r^{\mathrm{med}}(\omega)]^{\frac{3}{2}}\rbrace\right).
    \stepcounter{equation}\tag{\theequation} 
\end{align*}
As mentioned above, we have introduced the dimensionless scaling parameter $Y$ as $\gamma_{\mathrm{med}}(\omega,T,Y)\equiv Y\gamma_{\mathrm{med}}(\omega,T)$ to adjust the noise intensity, which is chosen independently of the type of medium. One can see that varying the parameter $Y$ effectively scales the magnitude of the environmental magnetic field $\hat{\mathbf{B}}_{\mathrm{med}}(t)$ by a factor of $\sqrt{Y}$. In addition, $\epsilon_r^{\mathrm{med}}(\omega)$ represents the complex relative permittivity specific to the medium \cite{gabriel,WANG_cole}, the parameter $T$ denotes the temperature of the environmental heat bath and $R$ is the cutoff wavelength of the environmental electromagnetic field. Since we focus on molecular systems, we set $R=1 $ \AA, which corresponds to a typical molecular size. The constants $\mu_0$ and $c$ are the magnetic constant and the speed of light, respectively. The mean occupation number of the photons \cite{QB} in a heat bath $ N(\omega,T)$, is given by
\begin{align}
    N(\omega,T)=\frac{1}{\operatorname{exp}\left({\frac{\hbar\omega}{k_{\mathrm{B}}T}}\right)-1},
\end{align}
where $k_{\rm B}$ is the Boltzmann constant. For each frequency difference of the RP system $\omega$, the noise operator $\hat{A}_{i\alpha}(\omega)$ acts with a specific coupling coefficient $\gamma_{\mathrm{med}}(\omega,T,Y)$.

\subsection{\label{subsec:CC}Coupling constants}
In this subsection, we discuss how to set the parameters of the coupling constants in Eq. (\ref{eq:CC}) to reflect realistic environmental conditions in vivo.

We first consider the type of medium surrounding the RP. The photoreceptive proteins that form RP, such as cryptochromes, are found in the photoreceptor cells that make up the avian retina \cite{Luo2024}. In the celluar environment, these molecules are surrounded by numerous water molecules. Several studies have investigated how various interactions between water molecules and the RP influence the mechanism of the chemical compass \cite{Hamada,WaterMP}. Based on these findings, we consider water as the medium for the environmental magnetic field. 

In addition to water, we consider the vitreous humor and retina as possible media, which are part of the biological tissue of the eye \cite{retina,collagen}. This choice is motivated by the fact that the RP is located between the vitreous humor and the retina. The former is a gel-like tissue composed of approximately 90$\%$ water \cite{VH}, and the latter is a film-like tissue comprising photoreceptor cells. The subscript ``med" in Eq.~(\ref{eq:CC}) refers to the three types: retina, VH (vitreous humor) and water.

The frequency dependence of the complex permittivity $\epsilon_r^{\mathrm{med}}(\omega)$ is unique to each medium \cite{gabriel}. In this study, we use the values from the databases for the complex permittivity of the three aforementioned tissues \cite{ITIS, Italy}. However, for convenience, we used human body measurement data as a substitute for those of the avian retina and vitreous humor in the calculations.

The remaining parameter to be set in $\gamma_{\mathrm{med}}(\omega,T,Y)$ is the temperature $T$. Here, we set $T$ of the heat bath at $T=$ 313 K (40 ° C), which corresponds to the typical body temperature of birds \cite{birdTemperature}.

\begin{figure}[htbp]
    \centering
    \includegraphics[width=7.5cm,clip]{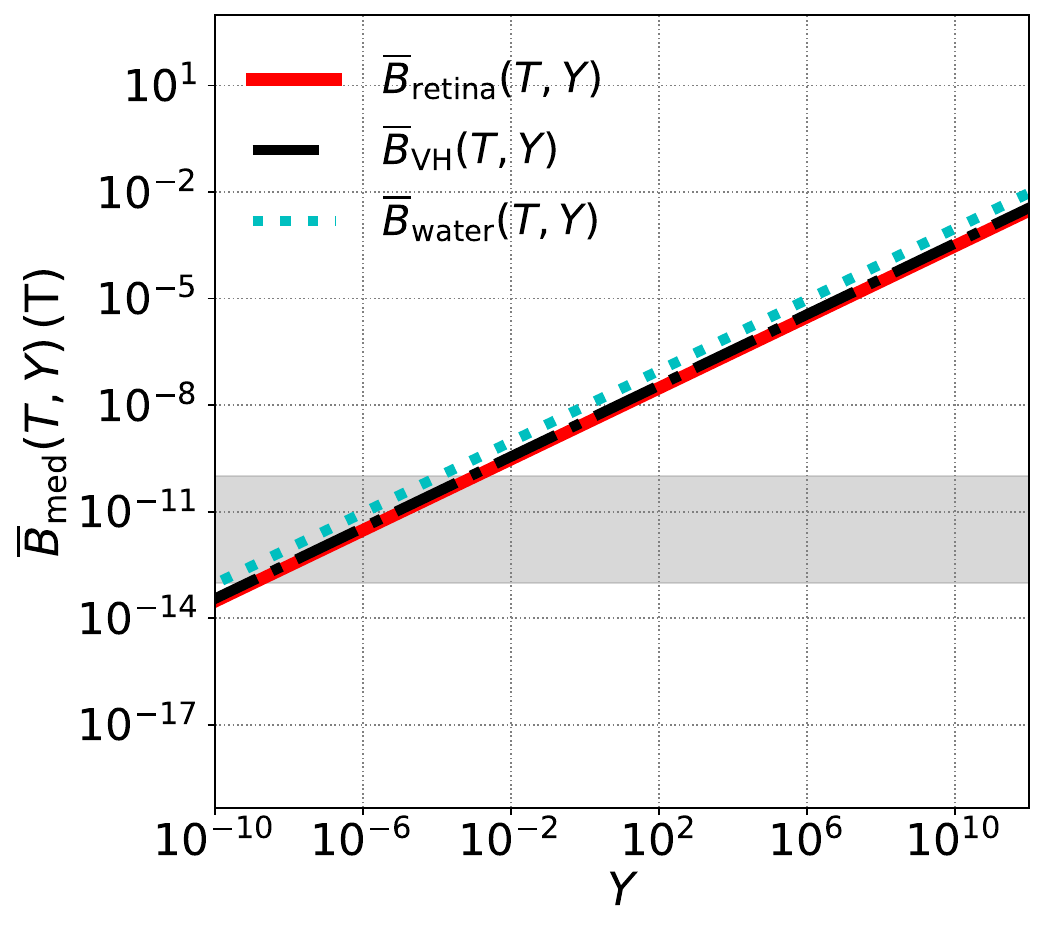}
    \caption{The $Y$-dependence of the mean magnetic field $\overline {B}_{\mathrm{med}}(T,Y)$. As described in the Sec. \ref{subsec:CC}, the temperature is fixed at $T=313$ K. The solid red and black dash-dot line represent the result for the retina and vitreous cases, respectively. The dotted light blue line represents the result for water as the medium in the cellular environment. The gray region indicates the biomagnetic range established in this study (\ref{eq:range}). }
    \label{fig:MFD}
    \vspace{-0.75em}
\end{figure}%

\begin{figure*}[htbp]
    \centering
    \includegraphics[width=17.5cm,clip]{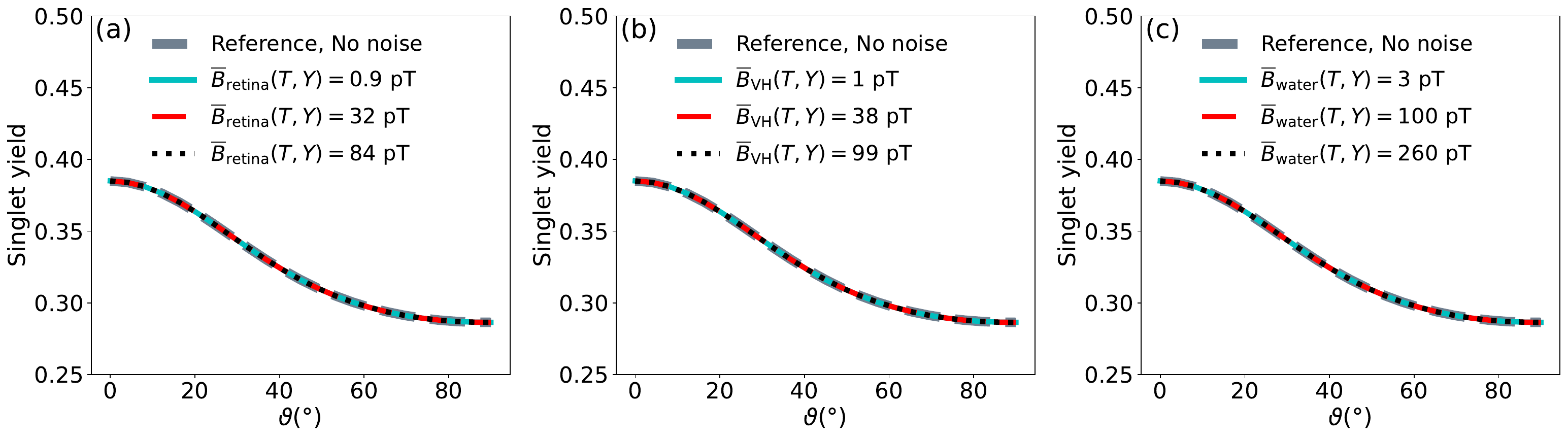}
    \caption{The angular dependence of the singlet yield in the radical pair model under environmental noise caused by magnetic fields within the biomagnetic range (\ref{eq:range}). Panels (a), (b), and (c) shows the results for the retina, vitreous humor, and water as the medium, respectively. In each case, $Y$ values of $1.00\times10^{-7}$, $1.15\times10^{-4}$, and $7.76\times10^{-4}$ were used to adjust the strength of the environmental magnetic field. No noise effects were observed, and the ideal curve that enables the unique determination of inclination angle $\vartheta$ from the yield was preserved.}
    \label{fig:c1}
    \vspace{-0.75em}
\end{figure*}

\subsection{\label{subsec:Biomagnetism}Biomagnetic range}

In this subsection, we estimate the magnitude of the environmental magnetic field.As shown in Eqs. (\ref{eq:Gamma1}) and (\ref{eq:Gamma2}), the coupling strength $\gamma_{\rm{med}}(\omega, T)$ is expressed in terms of the autocorrelation function of the magnetic field. According to Parseval’s theorem, the following relation holds between the time and frequency domains of the Fourier transform:
\begin{align}
     \frac{1}{2 \pi}\int_{-\infty}^{\infty}\left|\frac{\hbar^2}{\mu_{\mathrm{B}}^2}\gamma_{\mathrm{med}}(\omega,T)\right|^2 d\omega=\int_{-\infty}^{\infty}|\Gamma(t)|^2 dt\equiv || \Gamma(t) ||_2.
\end{align}
Thus, we can estimate the mean value of the environmental magnetic field as the square root of $l_2$-norm of $\Gamma(t)$:
\begin{align*}\label{Parseval}
    \overline {B}_{\mathrm{med}}(T) &\equiv \sqrt{|| \Gamma(t) ||_2}\\
    &=\Big[\frac{1}{2 \pi}\int_{-\infty}^{\infty}\left|\frac{\hbar^2}{\mu_{\mathrm{B}}^2}\gamma_{\mathrm{med}}(\omega,T)\right|^2d\omega\Big]^{\frac{1}{4}}.
    \stepcounter{equation}\tag{\theequation} 
\end{align*}
When the scaling parameter $Y$ is introduced and
the coupling constant $\gamma_{\mathrm{med}}(\omega, T ,Y)$ is used, the
corresponding magnetic field strength effectively becomes $ \overline {B}_{\mathrm{med}}(T,Y)=\sqrt{Y}\times\overline {B}_{\mathrm{med}}(T)$. 
As shown in this formula, the mean magnetic field is determined by the parameters related to the biological environment, which are encapsulated in $\gamma_{\mathrm{med}}(\omega,T,Y)$ and depends on the type of medium. Figure \ref{fig:MFD} shows the mean magnetic field $\overline {B}_{\mathrm{med}}(T,Y)$ as a function of $Y$. As seen in this figure, when $Y=1$ is given, the magnetic field strength $\overline {B}_{\mathrm{med}}(T,Y)$ is on the order of nano teslas, regardless of the medium. This represents the theoretical value for the contribution from thermal fluctuations of the electromagnetic field in the medium. Based on this result, we can derive the relation between the parameters of the RP model and the environmental magnetic field, providing a link between the theoretical model and the realistic in vivo environments.

The most important factor in our discussion is the magnitude of the environmental magnetic field. We consider biomagnetism as the naturally occurring environmental magnetic field. Biomagnetism is extremely subtle and it is known that measuring it, especially in small animals such as migratory birds, is very difficult. For example, the magnetic field of the human heart is about 50 to 100 pT, while that of the brain is on the order of 100 fT \cite{humanmag}. Even in measurements using animals such as guinea pigs, the heart's magnetic field is reported to be on the order of tens of picotesla \cite{guinea}, and the biomagnetic field around the eyes of migratory birds is similarly estimated to be at most tens of picotesla or less. 
Therefore, we set the upper limit of magnetic field strength that may exist in a natural biological magnetic environment to 100 pT.
Here, the range of the mean environmental magnetic field
\begin{align}\label{eq:range}
    \overline {B}_{\mathrm{med}}(T,Y) \leq 100\,\mathrm{pT},
\end{align} 
corresponds to the typical magnetic fields for biomagnetism \cite{williams1977,humanmag}. In this range, our calculations only consider noise derived from biomagnetism. By varying the angle of geomagnetic inclination $\vartheta$ and solving the master equations \cite{vectorize}, we obtain the population of the singlet shelving state $|S\rangle$ (i.e., the singlet yield values). This enables us to assess the angular dependence of the singlet yield. In the first part of the next section, we will adjust the value of $Y$ so that the magnetic field strength $ \overline {B}_{\mathrm{med}}(T,Y)$ remains within the range $ \overline {B}_{\mathrm{med}}(T,Y)\leq100$ pT, and investigate the effects of noise from the biomagnetic field. In the latter part, we will investigate the effects of strong magnetic field noise exceeding 100 pT by adjusting the value of $Y$.

\begin{figure*}[htbp]
    \centering
    \includegraphics[width=17.5cm,clip]{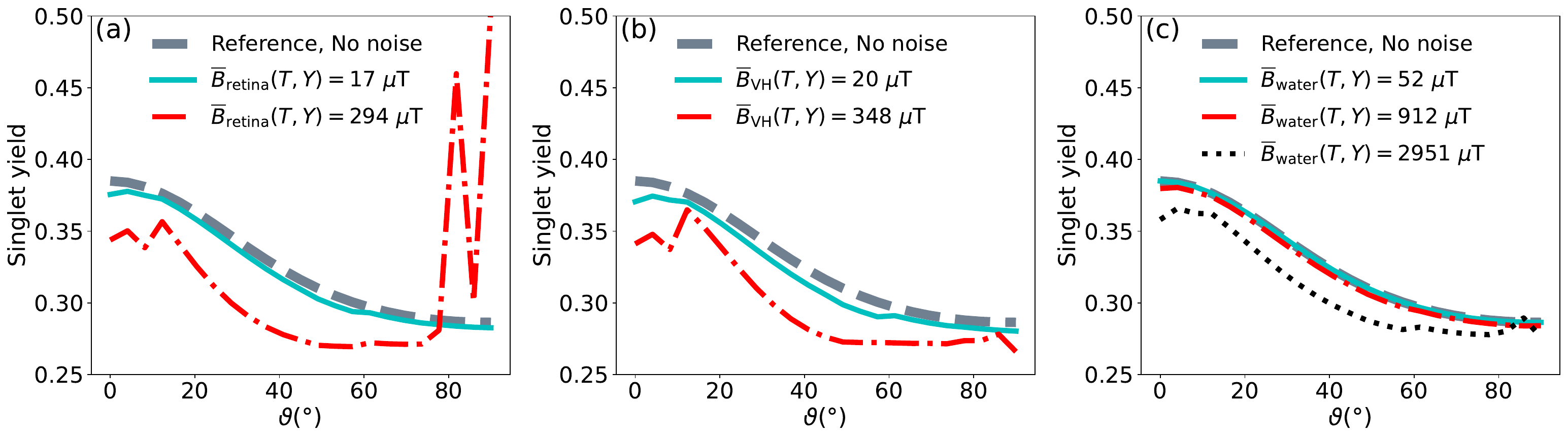}
    \caption{The angular dependence of the singlet yield in the radical pair model under environmental noise caused by magnetic fields. Panels (a), (b), and (c) show the results for the retina, vitreous humor, and water as the medium, respectively. In each case, $Y$ values of $3.16\times10^{7}$, $9.55\times10^{9}$, and $1.00\times10^{11}$ were used to adjust the strength of the environmental magnetic field. Noise effects were more pronounced in biological tissues, such as the retina and vitreous humor, compared to water.}
    \label{fig:c2}
    \vspace{-0.75em}
\end{figure*}%

\section{\label{sec:result}Results}
First, we present the angular dependence of the singlet yield obtained from numerical calculations in the picotesla range, including the upper limit of the biomagnetic range at 100 pT (\ref{eq:range}).

Figure \ref{fig:c1}(a) shows the results for the case where the medium for the environmental magnetic field is the retina. The parameter $Y$ is adjusted so that the corresponding magnetic field falls within the biomagnetic range (\ref{eq:range}). The values of $Y$ for the light blue, red, and black plots are $1.00\times10^{-7}$, $1.15\times10^{-4}$, and $7.76\times10^{-4}$, respectively. The three lines almost overlap with the result without environmental noise (shown in dashed gray line), indicating the noise has no significant effect. Even near the upper limit of the biomagnetic range (100 pT), the yield curve depends monotonically on the inclination angle $\vartheta$, maintaining the compass function. This implies that $\vartheta$ can be uniquely determined from the yield. Figures \ref{fig:c1} (b) and \ref{fig:c1} (c) show the results for vitreous humor and water as the medium, respectively, using the same $Y$ values as in Fig. \ref{fig:c1}(a). In all cases, the plots confirm that the noise has no noticeable effect. Based on these results, we can conclude that the chemical compass described by the RP model is highly robust against biomagnetic noise, with no disturbance or disorientation of its functionality.

Next, we investigate how the compass function is disrupted when the noise strength is increased. The noise effect appeared only after the parameter $Y$ exceeded 1 and increased by several million times.

Figures \ref{fig:c2} (a) and \ref{fig:c2} (b) show the results for the retina and vitreous humor as medium, respectively. The values of $Y$ for the light blue and red plots are $3.16\times10^{7}$ and $9.55\times10^{9}$ respectively, corresponding to environmental magnetic fields ranging from tens to hundreds of microteslas. Unlike the previous results within the biomagnetic range, the yield curves in these cases exhibit significant deformation, including reduced contrast and loss of monotonicity, indicating that the compass function is no longer maintained. Figure \ref{fig:c2}(c) shows the results for water as the medium. In addition to the case of $Y=3.16\times10^{7}$ and $9.55\times10^{9}$, the case of $Y=1.00\times10^{11}$ (black plot) is also included, where the magnetic field of the environmental magnetic field exceeds several thousand microteslas. Although the effects are less pronounced than the retina or vitreous humor, the yield curve still shows deformation, reduced contrast, and a loss of monotonicity. These results demonstrate that the compass described by the RP model is significantly disrupted under environmental magnetic noise at field strength of hundreds to thousands of micro teslas. However, these values are far beyond the biomagnetic range (\ref{eq:range}). Such high magnetic field strengths are rare in natural environments, but can occur due to artificial devices such as electric motors and generators \cite{Nishimura, WHO}.

The objective of this study is to bridge discussions on the avian chemical compass, a quantum mechanism operating in a biological environment, with the magnitude of environmental magnetic field, particularly in relation to the potential influence of biomagnetism. As demonstrated by the results,  model calculations suggest that the avian compass described by the RP model is robust against biomagnetic noise. 

\section{\label{sec:Discussion}Discussion}
In this section, we assess the limitations and validity of this study in the context of the actual structure of RP and biological environments.

First, as seen in (\ref{eq:hamitonian}), this study is based on a simple toy model. Furthermore, various interactions such as exchange and dipolar interactions between radicals \cite{scavengers} are assumed to be negligible and are therefore excluded from our model. 

Given the structural complexity of the protein molecules proposed as candidates for the RP model \cite{PRL,relativeorientation,Luo2024}, a more realistic approach would require a spin Hamiltonian that includes at least several dozen nuclear spins that form a spin bath \cite{relativeorientation, withoutCoherence}, coupled to unpaired electron pairs through HFI.

\begin{figure}[htbp]
    \centering
    \includegraphics[width=7.5cm,clip]{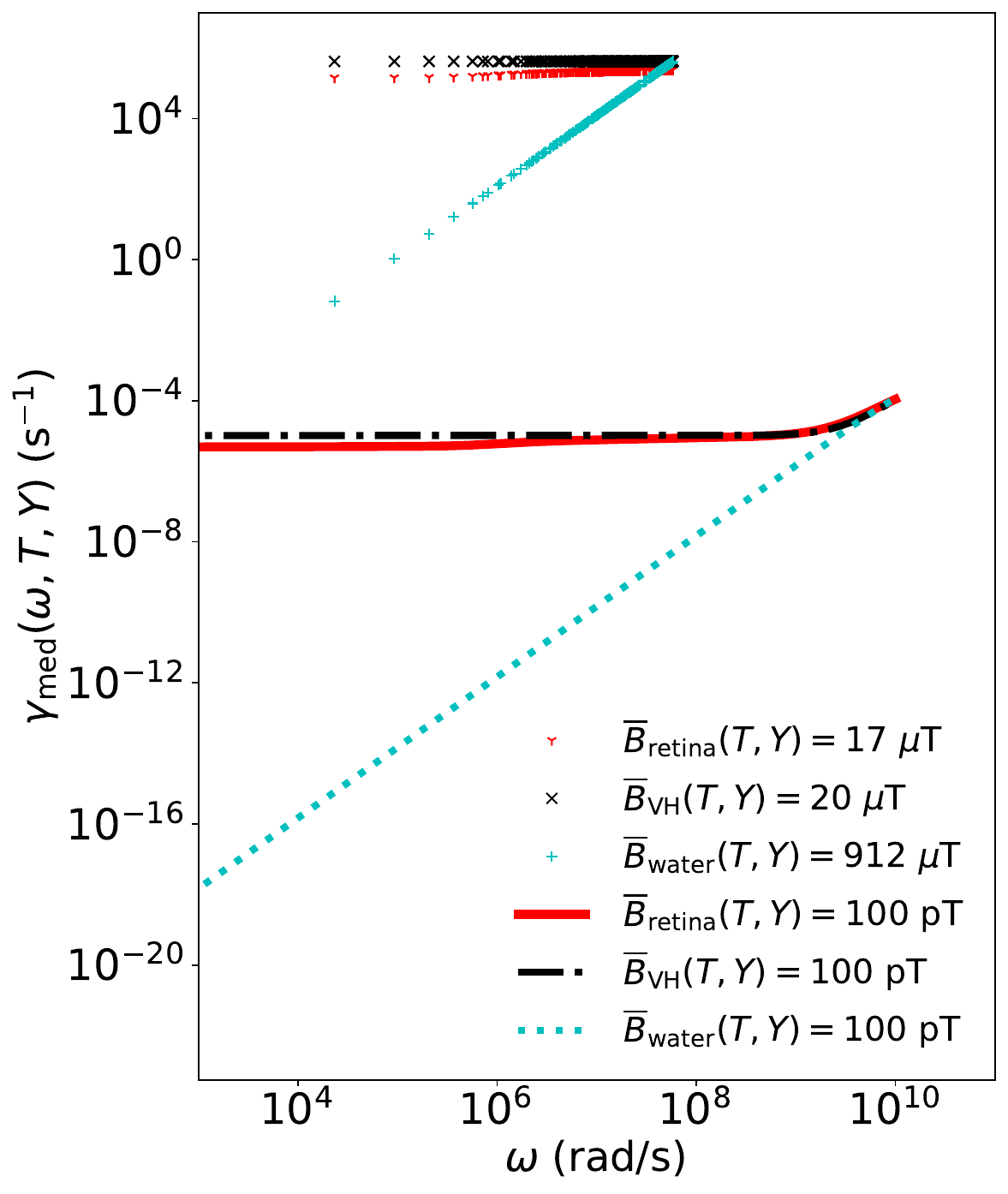}
    \caption{Frequency $\omega$ dependence of the coupling constant $\gamma_{\mathrm{med}}(\omega,T,Y)$. The dots represents the values of the coupling constant that appeared in the calculation shown in Fig.~\ref{fig:c2}, where noise effects on the singlet yield curve were observed. The lines indicate the coupling constant values at the upper limit of the biomagnetic range (\ref{eq:range}) 100 pT, as shown in the legend.}
    \label{fig:CC}
    \vspace{-0.75em}
\end{figure}%

In this paper, the coupling constant between the RP system and the environment is determined by the total magnitude of the coupling constant (\ref{eq:CC}). This magnitude is influenced not only by the biological environmental properties but also significantly by the transition frequencies $\omega$ between the eigenstates of the system Hamiltonian (\ref{eq:hamitonian}). In the toy model used here, the transition frequencies of the spin Hamiltonian range from kHz to several hundred MHz. In contrast, a more complex spin Hamiltonian with a larger number of spins will likely have an expanded energy eigenvalue distribution, with large transition frequencies included. Consequently, even if the environmental magnetic field has the same magnetic field, the coupling constant may vary. Figure \ref{fig:CC} shows that the coupling constant $\gamma_{\mathrm{med}}(\omega,T,Y)$ tends to increase as the frequency order $\omega$ increases. However, the scale of coupling constant required for noise to impact the yield curve, as illustrated by dots in Fig. \ref{fig:CC}, remains quite different from these values at the upper limit of the biomagnetic range (\ref{eq:range}). Thus, theoretically, it is unlikely that the coupling constant would reach levels sufficient to induce noise effects, even as the system size increases with the environmental magnetic field within the biomagnetic range.

However, as the number of spins and degrees of freedom of the system increase, the noise operators also grows exponentially. For a system with $d$ $S=1/2$ spins, the number of Lindblad operators (\ref{eq:LO}) per Pauli operator is given by $2^d(2^d-1)$. For example, an RP Hamiltonian with a spin bath containing 21 nuclear spins ($S=1/2$) and unpaired electrons \cite{relativeorientation} would require the implementation of roughly 70 trillion noise operators per Pauli operator. In even larger system, an immense number of noise operators must be incorporated into the time evolution. This issue is not unique to the RP model but is a general challenge in quantum open systems. Moreover, as the number of noise operators increases, the system is likely to be more susceptible to environmental influences.

Taking all these factors into account, enlarging the model Hamiltonian to reflect a more realistic setup would enhance the influence of environmental noise. However, the robustness of the system against noise could also change. Ultimately, to precisely analyze the effects of biomagnetic noise on a realistic RP model while maintaining consistency with the model in this paper, solving the Lindblad equation for a large system would be necessary, potentially requiring novel approaches.

Moreover, while this study has focused solely on magnetic fields as the environmental factor, warm and wet biological environments likely introduce additional influences on the function of the chemical compass, such as a quantum thermal noise or generalized amplitude damping noise \cite{QB,thermal}. Furthermore, irregular factors such as geomagnetic disturbances caused by magnetic storms are not taken into account in this study. Developing an RP model that incorporates all such environmental factors could contribute to deeper insights on the relationship between environmental noise and system behavior, not only in quantum biology but across quantum open systems in general.

\section{\label{sec:summary}Summary}

We developed an RP model that incorporates environmental noise from environmental magnetic fields in various media. The media include water, which is representative of the cellular environment surrounding the RP, as well as the biological tissues that make up the eye, such as the retina and vitreous humor. This model was used to investigate the impact of environmental noise on the avian chemical compass described by the RP model. In all media, no impact from noise due to environmental magnetic fields within the biomagnetic range (i.e., magnetic fields on the order of tens to hundreds of pico teslas) was observed. However, noise from environmental magnetic fields with higher magnetic fields, in the range of tens of micro teslas to milli teslas, showed significant effects. Such a strong biomagnetic field is unlikely, but artificial magnetic field of this strength could exist and disrupet the compass function. Based on these results, we conclude that within our model calculations, the avian compass function described by the RP model is robust against the biomagnetic noise, but could be disrupted by artificial magnetic fields.

\nocite{}
\bibliography{apssamp}% Produces the bibliography via BibTeX.

\end{document}